\documentclass[aps,prl,reprint,superscriptaddress,twocolumn,showkeys]{revtex4-1}
\usepackage{graphicx}
\usepackage[draft]{hyperref}
\usepackage{amsmath}
\usepackage{siunitx}

\begin{document}

\title{Gaussian entanglement distribution with gigahertz bandwidth}

\author{Stefan Ast}
\email[]{stefan.ast@aei.mpg.de}
\affiliation{Institut f\"ur Laserphysik und Zentrum f\"ur Optische Quantentechnologien, Universit\"at Hamburg, Luruper Chaussee 149, 22761 Hamburg, Germany}
\affiliation{Institut f\"ur Gravitationsphysik, Leibniz Universit\"at Hannover and Max-Planck-Institut f\"ur Gravitationsphysik (Albert-Einstein-Institut), Callinstra{\ss}e 38, 30167 Hannover, Germany}
\author{Melanie Ast}
\affiliation{Institut f\"ur Laserphysik und Zentrum f\"ur Optische Quantentechnologien, Universit\"at Hamburg, Luruper Chaussee 149, 22761 Hamburg, Germany}
\affiliation{Institut f\"ur Gravitationsphysik, Leibniz Universit\"at Hannover and Max-Planck-Institut f\"ur Gravitationsphysik (Albert-Einstein-Institut), Callinstra{\ss}e 38, 30167 Hannover, Germany}
\author{Moritz Mehmet}
\affiliation{Institut f\"ur Gravitationsphysik, Leibniz Universit\"at Hannover and Max-Planck-Institut f\"ur Gravitationsphysik (Albert-Einstein-Institut), Callinstra{\ss}e 38, 30167 Hannover, Germany}
\author{Roman Schnabel}
\email[]{roman.schnabel@physnet.uni-hamburg.de}
\affiliation{Institut f\"ur Laserphysik und Zentrum f\"ur Optische Quantentechnologien, Universit\"at Hamburg, Luruper Chaussee 149, 22761 Hamburg, Germany}
\affiliation{Institut f\"ur Gravitationsphysik, Leibniz Universit\"at Hannover and Max-Planck-Institut f\"ur Gravitationsphysik (Albert-Einstein-Institut), Callinstra{\ss}e 38, 30167 Hannover, Germany}

\date{\today}

\begin{abstract}
The distribution of entanglement with Gaussian statistic can be used to generate a mathematically-proven secure key for quantum cryptography. The distributed secret key rate is limited by the  {entanglement strength, the entanglement bandwidth and the bandwidth of the photo-electric detectors}. The development of a  source for strongly, bi-partite entangled light with high bandwidth promises an increased measurement speed and a linear boost in the secure data rate. Here, we present the experimental realization of a Gaussian entanglement source with a bandwidth of more than 1.25\,GHz. The entanglement spectrum was measured with balanced homodyne detectors and was quantified via the inseparability criterion introduced by Duan and coworkers with a critical value of 4 below which entanglement is certified. Our measurements yielded an inseparability value of about 1.8 at a frequency of 300\,MHz to about 2.8 at 1.2\,GHz extending further to about 3.1 at 1.48\,GHz. In the experiment we used two 2.6\,mm long monolithic periodically poled potassium titanyl phosphate (PPKTP) resonators to generate two squeezed fields at the telecommunication wavelength of 1550\,nm.  Our result proves the possibility of generating and detecting strong continuous-variable entanglement with high speed.
\end{abstract}


\maketitle
\begin{figure}[tbp]
	\centering	\includegraphics[width=1.00\columnwidth]{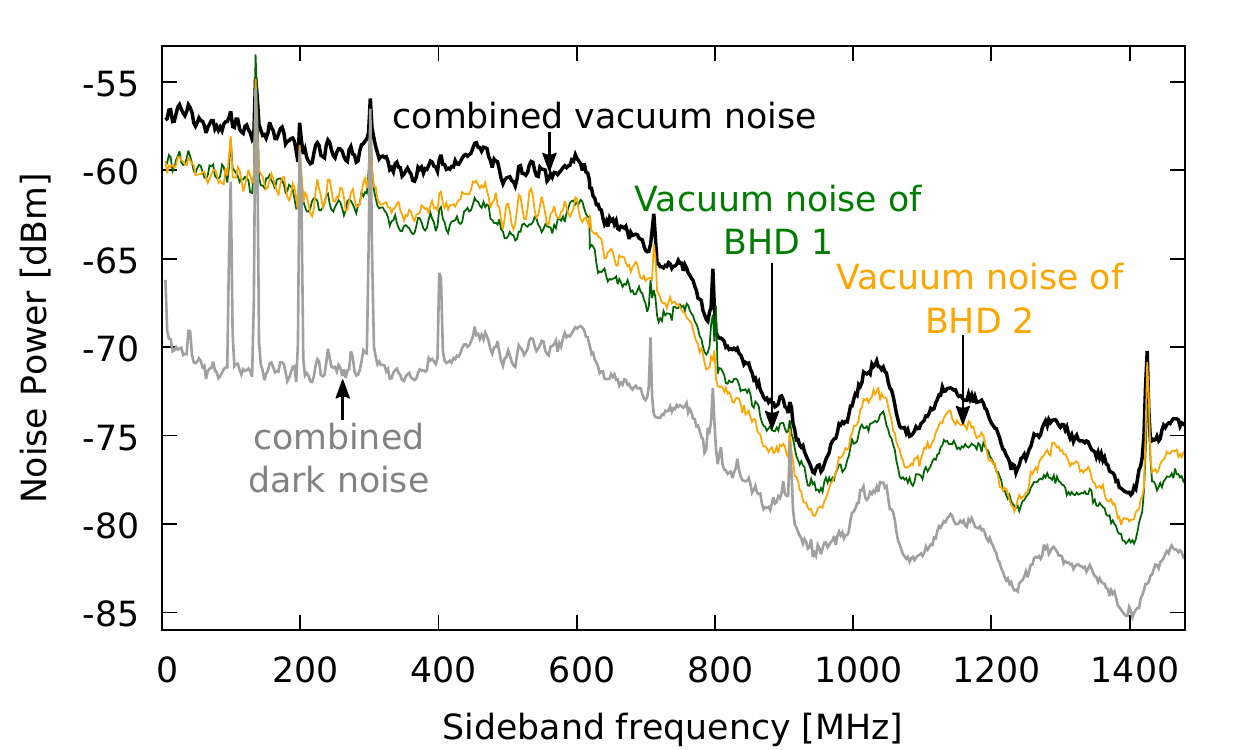}
	\caption[Vacuum noise levels for the subtraction of two homodyne detectors]{Characterization of the two GHz-bandwidth balanced homodyne detectors (BHDs). The plot shows the vacuum noise spectra of BHD\,1 (green) and of BHD\,2 (yellow) as well as the vacuum noise level for the combination of the signals (black). The dark-noise (gray) clearance for sum of both homodyne detectors is around 13\,dB at frequencies up to 300\,MHz, below 5\,dB at 900\,MHz and about 7\,dB from 1\,GHz to 1.5\,GHz. The spectra are composed of a low frequency measurement from 0.3-620\,MHz with a local oscillator power of 5\,mW on each detector and a high frequency measurement from 620--1480\,MHz with 6\,mW LO on BHD\,1 and 3\,mW LO on BHD\,2, respectively. The regions of differing vacuum noise levels for yellow and green result in a degradation of the measured entanglement strength at these frequencies. The measurements were performed with an RBW of 3\,MHz, a VBW of 1\,kHz and a sweep time of 540\,ms.}
	\label{fig:GHz-s-class-Ent-paper}  
\end{figure}
\begin{figure}[tb]
	\centering	\includegraphics[width=1.0\columnwidth]{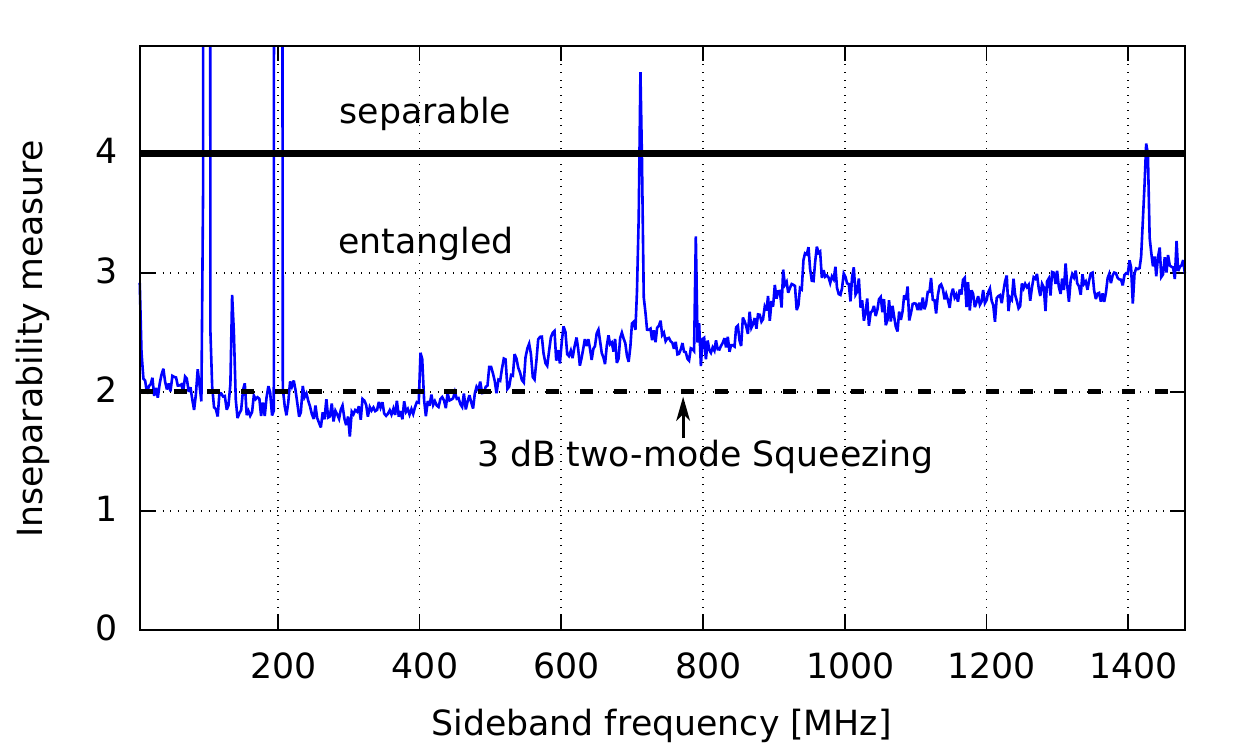}
	\caption[Normalized spectra of the s-class entanglement measurement]
	{Spectrum of the Gaussian entanglement measure $[\Delta^{2}(\hat{\text{X}}_{\text{A}}+\hat{\text{X}}_{\text{B}}) + \Delta^{2}(\hat{\text{Y}}_{\text{A}}-\hat{\text{Y}}_{\text{B}})]$ averaged twice and normalized to the combined vacuum noise level (solid line). Entanglement is certified below the critical value of 4, and thus, here, over the entire spectrum. For values below 2 (dashed line) the observation directly certifies Einstein-Podolsky-Rosen (EPR) entanglement, since more than 3\,dB two-mode squeezing was observed \cite{Bowen-Schnabel-Lam-Ralph-PRL,Bowen-Schnabel-Lam-Ralph-PRA}. With the reasonable assumptions that the optical loss did not increase towards higher frequencies, together with the also likely assumption that the influence of dark noise did not become significant, EPR entanglement was present over the full spectrum shown. The measurements were performed with an RBW of 3\,MHz, a VBW of 1\,kHz and a sweep time of 540\,ms. The origin of the electronic peaks and the bumps in the spectrum is discussed in the text.}
	\label{fig:Normalized-s-class-Spektrum-Xa-Xb-Pa+Pb-23-07-2014-averaged-thesis}
\end{figure}

Entanglement is often referred to as the most important signature of quantum physics. It is a useful tool for fundamental research and it is used as a resource for applications in quantum information experiments~\cite{Quantum-Info}. The most mature quantum information technology today is quantum key distribution (QKD)~\cite{Quantum-Crypto} using discrete variables~\cite{BB84} or continuous variables~\cite{QKD-cont-var}. 
In the  {Gaussian regime} of continuous-variables, entanglement based QKD assuming the most general quantum attacks can be realized via two-mode squeezed states of light \cite{Furrer-QKD-Two-mode-sqz,Tobi-QKD-Nature}. The two-mode squeezing is typically generated by overlapping two squeezed fields on a 50/50 beam splitter \cite{Furusawa1998,Bowen2003}. The outputs of the beam splitter are entangled and can be used to distribute a quantum key between two parties, often called Alice and Bob.\\
The strength of the generated squeezing determines the entanglement strength, which is one of the important properties influencing the achievable secret key rate for  {single-sided device independent} QKD \cite{Tobi-QKD-Nature}. Entanglement values of below 0.36 have been realized in recent years \cite{Eberle-Stable-control,Steini-Entanglement}, where a value of 4 marks the upper threshold for an inseparable (entangled) state via the criterion by Duan et al. \cite{Duan-criterion}. For two identical squeezed modes, the Duan criterion reads $\Delta^{2}(\hat{\text{X}}_{\text{A}}+\hat{\text{X}}_{\text{B}})$ + $\Delta^{2}(\hat{\text{Y}}_{\text{A}}-\hat{\text{Y}}_{\text{B}})$ < 4, where the $\Delta^{2}$ marks a variance of joined quadrature operators for amplitude $\hat{\text{X}}$ and phase $\hat{\text{Y}}$ at the two parties Alice (A) and Bob (B). -- 
Another important property is the bandwidth over which the entanglement is observed. It is given by the  {joint bandwidths of entanglement source and detector. High bandwidths are} important for proposals to entangle an array of micro cavities \cite{Illuminati-Entanglement-rep,Illuminati-PRA} as well as for schemes using Einstein-Podolsky-Rosen (EPR) entangled light \cite{EPR-Paper,Reid-EPR-Criterion} for  {efficient and high-speed} quantum key distribution \cite{Tobi-QKD-Nature}. 
For the implemented QKD in \cite{Tobi-QKD-Nature}, an increased  {bandwidth of entanglement distribution} offers the advantage of higher robustness against decoherence and a higher secret key rate.\\ 
Current state-of-the-art experiments achieve high squeezing factors of up to  {15}\,dB \cite{Vahlbruch2016}, but at the expense of a comparably low bandwidth in the order of 100\,MHz or less \cite{Mehmet2010}. 
In Ref.~\cite{Waveguide-Squeezing} two-mode entangled beams were generated using periodically poled lithium niobate waveguides. A squeeze factor of 0.76\,dB was observed over a bandwidth of 30\,MHz. An actual entanglement bandwidth of up to 10\,THz was inferred. Ref.~\cite{On-Chip-Sqz} reported the `on-chip' generation of intensity difference squeezing using a cavity with GHz bandwidth. Up to 1.7\,dB was observed for frequencies up to 5\,MHz.  
In the pulsed laser regime, recently, a two-mode squeeze factor of about 2\,dB, corresponding to a Duan-value of 2.52, was observed over a bandwidth of 200\,MHz \cite{Ent-pulsed-GHz}. None of these experiments actually observed squeezing or two-mode squeezing of gigahertz bandwidth. It is noted that the latter three concepts envision a large bandwidth for entanglement distribution but have not proved the ability to generate strongly squeezed states.  
Here, we present the table-top realization of Gaussian entanglement distribution with GHz bandwidth at the telecommunication wavelength of 1550\,nm as a potential resource for future fiber-based QKD.
 {Our experiment builds on the successful previous concept that is able to generate up to 15\,dB of squeezing \cite{Vahlbruch2016,Mehmet2010} and up to 10\,dB of two-mode squeezing \cite{Eberle-Stable-control}, but uses a cavity design of GHz bandwidth \cite{ast2013Mono}. The outputs of two squeezing resonators were overlapped on a balanced beam splitter and the outputs measured with two home-made GHz-bandwidth balanced homodyne detectors.}
\\ 
\begin{figure}[!b]
	\centering	\includegraphics[width=1.0\columnwidth]{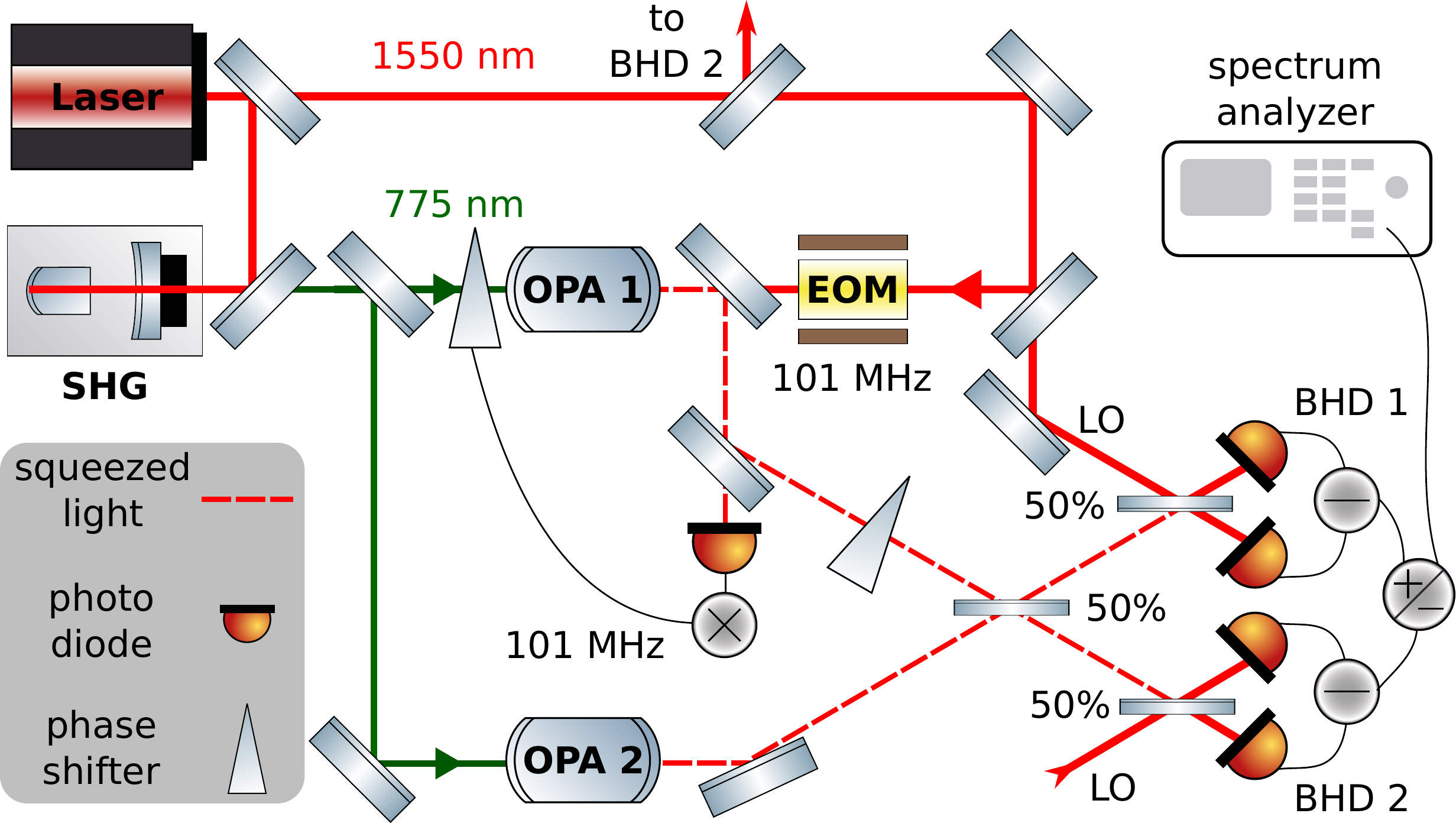}
	\caption[Setup for the GHz entanglement experiment]{Schematic setup for the experimental realization of high-bandwidth entanglement. Two monolithic PPKTP squeezing resonators of 2.6\,mm length generated squeezed light at 1550\,nm using optical parametric amplification (OPA\,1 \& OPA\,2). They were pumped by 775\,nm light coming from a single second harmonic generation cavity (SHG). The cavity lengths for the squeezing resonators were manually stabilized on resonance via temperature control. A control field was injected into OPA\,1 to stabilize the pump phase and the two homodyne detector (BHD\,1 \& BHD\,2) phases. The squeezed outputs of OPA\,1 and OPA\,2 were overlapped on a 50/50 beam splitter, thus creating the entangled beams. After subtracting and adding the two homodyne detector signals, respectively, a spectrum analyzer was used to acquire the field variances $\Delta^{2}(\hat{\text{X}}_{\text{A}}+\hat{\text{X}}_{\text{B}})$ and $\Delta^{2}(\hat{\text{Y}}_{\text{A}}-\hat{\text{Y}}_{\text{B}})$.}
	\label{fig:Monolith-GHz-Entangler-experimental-setup-locking-scheme-paper}
\end{figure}
The experimental setup consisted of an Ytterbium-doped fiber laser from \emph{NKT Photonics} type \emph{Koheras AdjustiK} including an external fiber amplifier type \emph{Koheras BoostiK} to generate light at 1550\,nm of 5\,W output power. The light first passed through a Faraday isolator to prevent back coupling of the laser light and then was transmitted through a mode-cleaner resonator to filter higher order modes and suppress high frequency amplitude and phase noise. A small fraction of about 10\,mW was used as a local oscillator for homodyne detection, while about 1.1\,W was frequency doubled inside a second harmonic generation cavity similar to the one used in~\cite{ast2011HCSHG}. 
The resulting 1\,W of 775\,nm laser light were used as pump fields for two squeezing resonators as schematically shown in Fig.~\ref{fig:Monolith-GHz-Entangler-experimental-setup-locking-scheme-paper}. Another mode-cleaner cavity (not shown in the picture for simplicity) filtered the frequency doubled field and acted as a diagnostic tool for mode-matching the 775\,nm pump light to the two squeezing resonators (OPA\,1, OPA\,2).\\
We used two monolithic crystals of the type presented in~\cite{ast2013Mono} for generating two single-mode squeezed beams. The squeezing resonators consisted of biconvex monolithic \mbox{PPKTP} crystals. The surfaces had radii of curvature of 12\,mm and reflective coatings for 775\,nm and 1550\,nm. The design values were $\text{R}_{1}= 99.98\,\%$, $\text{R}_{2}= 64\,\%$ at 1550\,nm and $\text{R}_{1}= 98\,\%$, $\text{R}_{2}= 99.98\,\%$ at 775\,nm, respectively. Together with the short crystal (cavity) length of 2.6\,mm and a refractive index of $\text{n}=1.816$ at 1550\,nm, this led to a resonator bandwidth of about 2.26\,GHz, a finesse of about 14 and a free spectral range of about 31.75\,GHz for the squeezed light wavelength of 1550\,nm. The finesse measured for the 775\,nm pump field was 307, corresponding to a resonant enhancement of the intra-cavity power by a factor of about 194. The squeezing resonator was pumped with up to 300\,mW of 775\,nm light (58\,W intra-cavity), which was limited by thermal effects in the nonlinear cavity. The maximal pump power was below the threshold power of 655\,mW (127\,W intra-cavity) for optical parametric oscillation. We numerically simulated the threshold by taking into account the intra-cavity waist sizes of $\text{w}_{0,1550}=\rm 33.86\,\si{\micro\metre}$ and $\text{w}_{0,775}=\rm 23.94\,\si{\micro\metre}$, the crystal's effective nonlinearity of $\text{d}_{\rm eff}=\rm 7.3\,\text{pm}/\text{V}$ as measured in \cite{ast2011HCSHG} and the crystal's absorption for both wavelengths. The absorption measurement for PPKTP yielded $\alpha_{1550}=84\pm 40\,\text{ppm}/\text{cm}$ and $\alpha_{775}=127\pm26\,\text{ppm}/\text{cm}$ for the fundamental and harmonic wavelengths, respectively, and were reported in ~\cite{{steinlechner2013absorption}}.\\
The squeezing cavities were manually tuned to resonate for the fundamental field close to their quasi-phase matching conditions by optimizing the crystal's temperatures via Peltier-elements. A phase-modulated control field at 1550\,nm in one of the resonators (OPA\,1) acted as a reference for stabilizing the measurement quadrature of both homodyne detectors to either the amplitude or the phase quadrature. Piezo-actuators in both local oscillator paths shifted the measurement phases for the homodyne detection. They are not shown in the picture for simplicity.\\
The entangled light was generated by overlapping the outputs of both squeezed light sources on a 50/50 beam splitter with a phase relation of 90$^\circ$. A piezo-actuated mirror adjusted the phase of one of the squeezed beams in the path before the beam splitter. A balanced homodyne detector in each output port of the beam splitter measured the two-mode squeezing, i.e.~the Gaussian entanglement. The two balanced homodyne detectors were identical, home-made and combined a large measuring bandwidth with highly efficient photodiodes. The latter were custom made InGaAs photodiodes with an active area of $60\,\si{\micro\metre}$. The homodyne detectors showed linear behavior over the full operation regime required for this work. Their electronic circuits used the directly subtracted currents of two photodiodes, where the outputs were divided into an AC and a DC path. The DC path used a transimpedance amplifier to convert the differential current into voltage. The AC path contained two microwave amplification transistors type \emph{Mini circuits MAR-6+} and \emph{Mini circuits ERA-5XSM+} for the high frequency AC signal generation. The AC outputs of the detectors were split up in two parts by power splitters from \emph{Mini circuits} type \emph{ZAPD-2-252-S+}. One part of the signal was used to extract error signals for locking the homodyne detector's phases. The other outputs of the power splitter's signals were subtracted/added with another \emph{Mini Circuits} power splitter type \emph{ZAPD-2-252-S+} over a frequency range of 0.3-1500\,MHz and acquired with a spectrum analyzer. This gave access to the field variances of the sum of the amplitude quadratures [$\Delta^{2}(\hat{\text{X}}_{\text{A}}+\hat{\text{X}}_{\text{B}})$] and the difference of the phase quadratures [$\Delta^{2}(\hat{\text{Y}}_{\text{A}}-\hat{\text{Y}}_{\text{B}})$] to verify and detect the entanglement.\\
The vacuum noise levels of the two homodyne detectors were found to be slightly different at high frequencies while using the same local oscillator (LO) power. This was caused by different amplifications of the electronic circuits, especially due to the slightly different gain spectrum of the microwave amplifiers. We could partly compensate for this effect by splitting the measurement into two bands and using an unbalanced setting of the LO powers for the measurement at high sideband frequencies. The vacuum noise spectra for both homodyne detectors in Fig.~\ref{fig:GHz-s-class-Ent-paper} were individually recorded from 0.3-620\,MHz with a local oscillator power of 5\,mW each and from 620--1480\,MHz with a local oscillator power of 6\,mW for detector 1 and 3\,mW for detector 2. The yellow and green lines in Fig.~\ref{fig:GHz-s-class-Ent-paper} show the vacuum noise levels of both individual homodyne detectors. Gray shows the combined dark noise level of the sum of both detectors and black the combined vacuum noise level of the sum of both detectors. The vacuum noises of both individual detectors were still at slightly different noise powers over large parts of the spectrum, especially above 500\,MHz sideband frequency. This presumably lead to a small degradation in the measured entanglement. The detector's combined dark noise clearance was about 13\,dB at 50\,MHz and was degrading to about 7\,dB at 1\,GHz. The finite bandwidth of the electronic circuit as well as the finite bandwidth of the photodiodes were the main contributions to the limited dark-noise clearance at high frequency. The 50/50 splitting ratio of the Mini Circuits power splitters used for extracting the locking signals were also decreasing the dark-noise clearance, but were necessary for pump phase and homodyne detection phase stabilization. Electronic pick-up was visible at a variety of frequencies of the homodyne detector's spectra and was mainly occurring due to Pound-Drever-Hall modulation frequencies at 101\,MHz, 138\,MHz as well as their harmonics. The peaks at 714\,MHz and 1428\,MHz were arising due to the free spectral range of the mode-cleaner cavities and are a result from the imperfect common mode rejection ratio of the homodyne detection. The photodiodes were sensitive for electronic pick-up noise due to a large TO housing as well as a combined pin for cathode and ground. They were effectively forming antennas, which contaminated large parts of the spectrum with electro-magnetic interspersals. However, we were able to shield the housing of the homodyne detector with commercially available microwave absorbing material to suppress some parts of the spurious electronic pick-up.%
\\
The measurement of the entanglement spectrum at 1550\,nm normalized to the vacuum noise level is plotted in Fig.~\ref{fig:Normalized-s-class-Spektrum-Xa-Xb-Pa+Pb-23-07-2014-averaged-thesis}. By detecting the variances of the sum of the amplitude quadratures [$\Delta^{2}(\hat{\text{X}}_{\text{A}}+\hat{\text{X}}_{\text{B}})$] and the variance of the difference of the phase quadratures [$\Delta^{2}(\hat{\text{Y}}_{\text{A}}-\hat{\text{Y}}_{\text{B}})$] the entanglement was observed as a non-classical noise suppression below the combined vacuum noise.
 We measured a non-classical noise suppression in the frequency band from 1--1480\,MHz for both the amplitude and the phase quadrature. The measurements of the two orthogonal quadratures were transferred to a Duan inseparability value, which is depicted in Fig.~\ref{fig:Normalized-s-class-Spektrum-Xa-Xb-Pa+Pb-23-07-2014-averaged-thesis}. They show EPR entanglement from 100--500\,MHz (below dashed line) with a minimal value of 1.8 at 300\,MHz and clearly demonstrate broadband entanglement from about 1--1480\,MHz sideband frequency (below solid line). The spectrum degrades to a value of about 3.1 at 1480\,MHz due to the squeezing resonator's finite linewidths as well as the homodyne detector's finite bandwidth. The noise curves recorded for the spectrum were averaged using two consecutive measurements. An inseparability value of below 2 (dashed line) is equivalent to individual squeezing values of slightly more than 3\,dB, which corresponds to a total detection efficiency of more than 50\%. Note that the total detection efficiency above 50\% marks the upper threshold for EPR entanglement \cite{Bowen-Schnabel-Lam-Ralph-PRL}. Since the detection efficiency should not degrade towards higher frequencies and the dark-noise clearance was almost 7\,dB up to 1.5\,GHz, EPR entanglement was probably present at all frequencies shown. The measured entanglement strength was limited by the total optical loss, the unequal vacuum noise levels due to the homodyne detector's gain and by the achieved pump power used in the experiment (300\,mW), which was far below the threshold for parametric oscillation at 655\,mW. The peaks showing up in the measurement below 300\,MHz are mainly due to the control field's phase modulations at 101\,MHz and 138\,MHz as well as their harmonics, which were used to stabilize mode-cleaner cavities, the pump phase of OPA\,1 and the homodyne detection phases.\\
We deduced the total optical detection efficiency in the experiment to be about 59\% from individual squeezing measurements. It consisted of the imperfect mode overlap at the homodyne detection beam splitters as well as the entanglement beam splitter (about 68\% efficiency), the transmission loss in the optical path of about 8\% (92\% efficiency) and the quantum efficiency of the photodiodes, which was deduced to be about $94\pm3\%$ from our previous experiments measuring squeezed light. 
The imperfect mode overlaps at the beam splitters gave the highest loss contributions. To align the mode matchings of the squeezed vacuum fields, auxiliary control beams were used. Their alignments to the squeezing resonators, however, were difficult, since the lengths of these resonators could not be scanned. Furthermore, we had to mode match the control beams in reflection of the squeezing resonators since they were highly under-coupled from their backs. 
The optical path transmission loss consisted mainly of imperfect HR coatings of the steering mirrors, imperfect AR coatings of lenses as well as 1\% mirror transmissions in the detection path used for tapping off locking signals for the phase stabilizations (see Fig.~\ref{fig:Monolith-GHz-Entangler-experimental-setup-locking-scheme-paper}). 
Another loss source for the entanglement was the finite homodyne detector's dark-noise clearance, in particular at frequencies around 900\,MHz. 
Finally, above 600\,MHz the entanglement was notably degraded due to the unbalanced vacuum noise levels of the two homodyne detectors, which result in a non optimal cancelation of correlated or anti-correlated noise when taking the difference or sum of their signals, respectively.\\
In conclusion, the experimental results reported in this paper serve as a proof of principle for the distribution of Gaussian entanglement with GHz bandwidth. We generated bi-partite entangled states of light at 1550\,nm and detected the entanglement via two fast home-made balanced homodyne detectors. The measurements showed more than 3\,dB of two-mode squeezing at frequencies between 100--500\,MHz and generic entanglement over the sideband spectrum from 1--1480\,MHz. Our measurements are quantified by a minimal Duan inseparability value of about $1.8 < 4$, which is consistent with the loss sources in our setup. The measured entanglement over a bandwidth of more than 1.48\,GHz is, to the best of our knowledge, the highest bandwidth for an entangled state observed so far. Further optimization of the experimental setup will likely enable the detection of 10\,dB squeezing, corresponding to an entanglement value of 0.4 with GHz bandwidth as simulated in \cite{ast2013Mono}. These entanglement values will be suitable for one-sided device independent QKD as recently demonstrated by Gehring \emph{et.~al} \cite{Tobi-QKD-Nature}. Our experiment therefore marks a first step towards high-speed fiber-based quantum key distribution for urban networks. 
\begin{acknowledgments}
The authors acknowledge support from the International Max Planck Research School on Gravitational Wave Astronomy (IMPRS-GW). M.A. acknowledges support from the Research Training Group 1991 by the Deutsche Forschungsgemeinschaft (DFG). M.M. acknowledges support from the Centre of Excellence for Quantum Engineering and Space-Time Research QUEST of the Leibniz Universit\"at Hannover.
\end{acknowledgments}

\end{document}